\documentstyle[floats,aps]{revtex}

\parskip=\medskipamount
  \font\elevenmib=cmmib10 scaled 1095
  \font\tenmib=cmmib10
  \font\eightmib=cmmib10 scaled 800
  \font\sixmib=cmmib10 scaled 667
  \skewchar\elevenmib='177
  \newfam\mibfam
  \def\mib{\fam\mibfam\tenmib}
  \textfont\mibfam=\tenmib
  \scriptfont\mibfam=\eightmib
  \scriptscriptfont\mibfam=\sixmib
  \mathchardef\alpha="710B
  \mathchardef\beta="710C
  \mathchardef\gamma="710D
  \mathchardef\delta="710E
  \mathchardef\epsilon="710F
  \mathchardef\zeta="7110
  \mathchardef\eta="7111
  \mathchardef\theta="7112
  \mathchardef\kappa="7114
  \mathchardef\lambda="7115
  \mathchardef\mu="7116
  \mathchardef\nu="7117
  \mathchardef\xi="7118
  \mathchardef\pi="7119
  \mathchardef\rho="711A
  \mathchardef\sigma="711B
  \mathchardef\tau="711C
  \mathchardef\phi="711E
  \mathchardef\chi="711F
  \mathchardef\psi="7120
  \mathchardef\omega="7121
  \mathchardef\varepsilon="7122
  \mathchardef\vartheta="7123
  \mathchardef\varrho="7125
  \mathchardef\varphi="7127
    \def\physgreek{
    \mathchardef\Gamma="7100
    \mathchardef\Delta="7101
    \mathchardef\Theta="7102
    \mathchardef\Lambda="7103
    \mathchardef\Xi="7104
    \mathchardef\Pi="7105
    \mathchardef\Sigma="7106
    \mathchardef\Upsilon="7107
    \mathchardef\Phi="7108
    \mathchardef\Psi="7109
    \mathchardef\Omega="710A}

\def\etal{{\it et al.\/}}

\def\ie{{\it i.e.\/}}

\def\eg{{\it e.g.\/}}

\def\sss#1{{\scriptscriptstyle #1}}

\def\ssr#1{{\sss{\rm #1}}}

\def\ONE{{1\kern-0.60em 1}}
\def\dsl{\raise.15ex\hbox{$/$}\kern-.57em\hbox{$\partial$}}
\def\nsl{\raise.15ex\hbox{$/$}\kern-.57em\hbox{$\nabla$}}
\def\id{\raise.72ex\hbox{$-$}\kern-.85em\hbox{$d$}\,}
\def\gtwid{\,{\raise.3ex\hbox{$>$\kern-.75em\lower1ex\hbox{$\sim$}}}\,}
\def\ltwid{\,{\raise.3ex\hbox{$<$\kern-.75em\lower1ex\hbox{$\sim$}}}\,}
\def\undr{\raise.3ex\hbox{$\sim$\kern-.75em\lower1ex\hbox{$|\vec
x|\to\infty$}}}

\def\frac#1#2{{\textstyle{#1 \over #2}}}
\def\half{\frac{1}{2}}

\def\fourth{\frac{1}{4}}

\def\sbl{\left [}
\def\sbr{\right ]}
\def\({\left (}
\def\){\right )}

\def\uar{\uparrow}
\def\dar{\downarrow}

\def\zhat{{\hat{\bf z}}}

\def\cH{{\cal H}}

\def\cO{{\cal O}}

\def\bfI{{\mib I}}

\def\bfS{{\mib S}}

\def\bfc{{\mib c}}

\def\bfi{{\mib i}}

\def\bfq{{\mib q}}

\def\bfdelta{{\mib\delta}}

\def\bftau{{\mib\tau}}

\def\xhi{{\raise.35ex\hbox{$\chi$}}}

\def\rmA{{\rm A}}

\def\rmC{{\rm C}}

\def\rmF{{\rm F}}

\def\rmN{{\rm N}}

\def\rmc{{\rm c}}

\def\ket#1{{\,|\,#1\,\rangle\,}}
\def\bra#1{{\,\langle\,#1\,|\,}}

\def\yds#1{^\dagger_{#1}}
\def\nds#1{_{#1}^{\vphantom{\dagger}}}

\def\and{a^{\phantom\dagger}}

\gdef\journal#1, #2, #3, 1#4#5#6{               
    {\sl #1~}{\bf #2}, #3 (1#4#5#6)}            
\def\pr{\journal Phys. Rev., }

\def\prb{\journal Phys. Rev. B, }

\def\prl{\journal Phys. Rev. Lett., }

\def\nupb{\journal Nucl. Phys. B, }

\def\phyla{\journal Phys. Lett. A, }

\def\jmp{\journal J. Math. Phys., }

\physgreek

\begin{document}
\draft

\def\be{\begin{equation}}
\def\ee{\end{equation}}
\def\bea{\begin{eqnarray}}
\def\eea{\end{eqnarray}}
\def\Tc{T_{\rmc}}
\def\c60{C$_{60}$}
\def\bb#1{C$_{60}^{#1}$}
\def\a3c60{A$_3$C$_{60}$}
\def\cHt{{\tilde\cH}}
\def\bdpar{\bfdelta_\parallel}
\def\bdper{\bfdelta_\perp}
\def\tpar{t^\parallel}
\def\tper{t^\perp}
\def\ubar{{\bar u}}
\def\alphabar{{\bar\alpha}}
\def\betabar{{\bar\beta}}
\def\kF{k_\ssr{F}}
\def\mubar{{\bar\mu}}

\twocolumn[\hsize\textwidth\columnwidth\hsize\csname @twocolumnfalse\endcsname
\title{TDAE-\c60: Can a Mott insulator be a ferromagnet?}
\author{Assa Auerbach and Daniel P. Arovas}
\address{Physics Department, Technion-IIT, Haifa 32000, Israel}


\date{\today}

\maketitle

\begin{abstract}
Motivated by the structure of TDAE-\c60, we  derive a multicomponent
superexchange Hamiltonian for the spins and orbital (``isospin'')
degrees of freedom in
the Mott-insulating phase.  We explore its phase diagram and identify
points of special interest: an SU(4) antiferromagnet (solved by
Sutherland) and two ferromagnet$\otimes$antiferromagnet points where the
ground state is known.  For the ferromagnetic regime, we apply interchain
mean field theory and derive an expression for the Curie temperature
where spin ordering occurs and a lower Ne{\'e}l temperature for a conjectured
isospin ordering.

\end{abstract}

\pacs{PACS numbers: 33.10.Lb,71.38.+i,74.20.-z,71.10.+x }

\vskip2pc]

\narrowtext

The fullerene compound TDAE-\c60, where \c60 is buckminsterfullerene
and TDAE is tetrakis\-(dimethylamino)\-ethylene C$_2$N$_4$\-(CH$_3$)$_8$,
exhibits ferromagnetism at $\Tc \approx 16^\circ$K \cite{tdae-fm,tdae-fm1}.
The striking aspects of this discovery are (i) the magnitude of $\Tc$ --
relatively large for a material with no transition metals --
and (ii) its nonmetallic conductivity, suggestive of Mott-Hubbard
localization\cite{microwave}.

ESR studies \cite{tanaka} show that TDAE donates an electron to \c60.
Furthermore, no ESR signature of TDAE$^+$ is observed, suggesting that
the TDAE radical spins are somehow paired.  The monoclinic structure
makes for a relatively short inter-\c60 separation along the $c$-axis.
We are then led to consider a model of \bb{1} chains whose conduction
electrons interact via superexchange.

Superexchange in a one-band model is always antiferromagnetic,
since the intermediate state $\ket{\uar\dar}$ is a spin singlet.
The molecular degeneracy of the $t_{1u}$ \c60 LUMO leads
to interesting possibilities not realized in a orbitally nondegenerate
model.  Indeed, Seshadri \etal \cite{rao1} have discussed how
ferromagnetism naturally arises via superexchange through intermediate
states with a negative singlet-triplet splitting (\eg\ Hund's rule)
in \bb{2-}.  Motivated by Ref. \cite{rao1}, we introduce here what
we believe to be a `minimal model', based on the structure of
TDAE-\c60, which leads to insulating ferromagnetic behavior.

The main result of our Letter is the full multicomponent superexchange
Hamiltonian for the Mott insulator, an analysis of its phase diagram
and ground states, and a discussion of three-dimensional
ordering in this quasi-one-dimensional system.  The model is
characterized by three coupling constants, which represent the
superexchange interaction strength through the three different
intermediate states of \bb{2-}.  In the limit where all the
intermediate states are degenerate, we obtain
an SU(4) Heisenberg antiferromagnet in its fundamental
representation, solved by Sutherland\cite{suth} using Bethe's
{\it Ansatz}.  In the case of large singlet-triplet splitting,
we are able to prove using Marshall's theorem \cite{marshall}
that the ground state is fully polarized in the spin variables and
is given by Bethe's wavefunction for the $S=\half$ antiferromagnet
in the isospin variables.  This latter case, which we think relevant
to TDAE-\c60, is a rare example where ferromagnetism can be proven in
a microscopic model of localized electrons \cite{comm1}.
The three-dimensional
susceptibilites can be expressed in terms of the exact one-dimensional
functions following Scalapino \etal \cite{sip}.
For the ferromagnetic model we find the spin Curie temperature and the orbital
N\'eel temperature as a function of interchain couplings.
Full details of the calculations below will be soon available in a longer paper
\cite{long}.

{\em The Hopping Model:}
We consider tight binding hopping on a lattice of \c60 molecules
with a filling of one electron per site.
In general, a tetragonal or monoclinic
crystalline symmetry will resolve the triply degenerate $t_{1u}$ orbital
into three distinct levels. Better details could be obtained
{\em ab-initio} once the precise structure
and orientations of the \c60 molecules are experimentally ascertained.
In our model, we shall retain only
what we believe may be the essential microscopic physics undelying
the ferromagnetism in TDAE-\c60:
(a) The hopping is quasi-one dimensional along the $c$-axis.
(b) We assume that the crystal field resolves the
$t_{1u}$ orbital triplet into a lower doublet ($l=\pm$) and
a higher singlet $l=0$ at higher energy, as if the crystal
fields are cylindrically symmetric about an axis which pierces the
center a pentagonal face of \c60.
(c) Hopping along the chains is assumed to preserve the orbital
magnetization $l$.

Thus we investigate the Hamiltonian
$\cH =\cH^\parallel_{\rm hop} + \cH_{\rm hop}^\perp+\cH_{\rm ion}$,
where
\bea
\cH_{\rm hop}^\parallel & = & -\tpar\sum_{\bfi,l,\sigma}\(c\yds{l\sigma}(\bfi)
c\nds{l\sigma}(\bfi+\bfc)+{\rm H.c.}\)\nonumber\\
\cH_{\rm hop}^\perp & = & -\half\sum_{\bfi,\bdper\atop l,l',\sigma}
\tper_{ll'}(\bdper)\( c\yds{l\sigma}(\bfi)c\nds{l'\sigma}(\bfi+\bdper)+
{\rm H.c.}\)\nonumber\\
\cH_{\rm ion} & = & \sum_{\bfi,\Lambda}\ubar_\Lambda\,\ket{\Lambda(\bfi)}
\bra{\Lambda(\bfi)}\ .
\eea
Here $c\yds{l\sigma}(\bfi)$ creates, at site $\bfi$,
an electron of spin polarization $\sigma=\uparrow,\downarrow$ and
``isospin'' $l=+,-$. $\bfc$ and $\bdper$ denote
nearest neighbor lattice vectors in the $c$ direction and the
$a$-$b$ plane respectively. $t_{ll'}$ are the hopping matrix elements between
orbitals $l$ and $l'$ on  neighboring chains.

$\cH_{\rm ion}$ is the interaction Hamiltonian which
discourages multiple electron occupancy on any \c60 molecule.
It is parametrized by pseudopotentials $\ubar_\Lambda$ which correspond to the
following  \bb{2-} multiplets,
\bea
\ubar_0:&& \frac{1}{\sqrt{2}}(c\yds{+\uar}c\yds{-\dar}\!-\!c\yds{+\dar}
c\yds{-\uar})
\!\ket{0}\nonumber\\
\ubar_1:&&
c\yds{+\uar}c\yds{-\uar}\ket{0},
\frac{1}{\sqrt{2}}(c\yds{+\uar}c\yds{-\dar}+c\yds{+\dar}c\yds{-\uar})
\ket{0} ,
c\yds{+\dar}c\yds{-\dar}\ket{0}\nonumber\\
\ubar_2:&&
c\yds{+\uar}c\yds{+\dar}\ket{0},
c\yds{-\uar}c\yds{-\dar}\ket{0}
\eea
The relations between $\ubar_\Lambda$ and the isotropic pseudopotentials $u_L$
of angular momenta $L$  are: $\ubar_1= u_1$, $\ubar_2=u_2$, but $\ubar_0=
{2\over 3} u_0 + {1\over 3} u_2$ due to projecting out the ``$l=0$''
orbital state.  Thus, while in an isotropic environment there might be pair
binding  ($u_0 < 0$) due to electron-electron \cite{ck} and electron-vibron
\cite{amt} interactions, it does not preclude
a repulsive $\ubar_0 > 0$ in the monoclinic crystal field environment.
This may help to explain why TDAE-\c60 is not a CDW, nor
a superconductor as is \a3c60.

{\em Multicomponent Superexchange Hamiltonian.}
Experiments have shown that TDAE-\c60 is insulating at low temperatures,
consistent with the existence of a gap to charge fluctuations
(i.e. all $\ubar_\Lambda>0$) \cite{microwave}.
The low-lying excitations can be described by a superexchange
Hamiltonian, formally obtained as a second order expansion in small
$\tpar/\ubar$. Since charge excitations are gapped, a renormalized
version of the superexchange Hamiltonian is expected to describe the low
energy excitations also for $\tpar/\ubar\gtwid 1$.

The zeroth order states of the superexchange Hamiltonian are four singly
occupied states enumerated by $l,\sigma$. The  operators
which act on these states can be represented by
spin  operators,
$S^\mu_\bfi=\half\sum_{l,\sigma,\sigma'} c\yds{l\sigma}(\bfi)\,
\tau^\mu_{\sigma\sigma'}\,c\nds{l\sigma'}(\bfi)$
and  ``isospin'' operators,
$I^\nu_\bfi=\half\sum_{l,l',\sigma} c\yds{l\sigma}(\bfi)\,
\tau^\nu_{ll'}\,c\nds{l'\sigma}(\bfi)\ $
where $\bftau$ are the Pauli matrices.
Taking into account the constraint
$\sum_\alpha c\yds{\alpha}c\nds{\alpha}=1$, the 15 independent elements of
the SU(4) generators
$S^\alpha_\beta= c\yds{\alpha}c\nds{\beta}$ can be expressed in terms of
the 15 operators
$\{S^\mu,I^\nu,S^\mu I^\nu\}$.

{\em One-Dimensional Limit.}
We first consider the purely one-dimensional case, where
$t^\perp_{ll'}=0$.  There are three superexchange constants defined as
\be
J_M\equiv {2(\tpar)^2\over\ubar_M}\ \,\qquad M=0,1,2\ ,
\ee
After a straigthforward calculation, we derive the
general spin/isospin representation of the effective Hamiltonian as
\bea
\cHt&=&\sum_n\(A\,\bfS_n\cdot\bfS_{n+1}+B\,\bfI_n\cdot\bfI_{n+1}
+C\, I^z_n I^z_{n+1}+\right.\label{sex1}\\
&&\left.+D\,\bfS_n\cdot\bfS_{n+1}\,\bfI_n\cdot\bfI_{n+1}
+E\,\bfS_n\cdot\bfS_{n+1}\,I^z_n I^z_{n+1} +F\)\ ,\nonumber
\eea
where
\bea
A =-\half J_1+J_2+\half J_0,&~~~~
&B=  \frac{3}{2} J_1-\half J_0,\nonumber\\
C=  J_0-J_2 ,&~~~~
&D =  2 J_1+2 J_0,\nonumber\\
E=  4 J_2 - 4 J_0 ,&~~~~
&F =  -\frac{3}{8} J_1 - \fourth J_2 - \frac{1}{8} J_0\ .
\eea
This model possesses a global SU(2)$\times$U(1) symmetry, \ie\ $\cHt$
commutes with $\sum_n \bfS_n$ and with $\sum_n I^z_n$.  Enlarged symmetries
occur when $J_0=J_2$, where the symmetry group is SU(2)$\times$SU(2), and
when $J_0=J_1=J_2$, where the symmetry group is SU(4).

{\em SU(4) Point --} At the point $\ubar_1=\ubar_2=\ubar_0\equiv\ubar$
(\ref{sex1}) acquires full SU(4) symmetry. For each $c$-chain the Hamiltonian
is
\be
\cH_{SU(4)}=J\sum_n \sum_{\alpha,\beta}S^\alpha_\beta(n)S^\beta_\alpha(n+1)
\ ,
\ee
where $J=2(\tpar)^2/\ubar$.

The SU($P$) Heisenberg antiferromagnet in the fundamental
representation has been solved by Sutherland for general $P$ using
Bethe's {\it Ansatz\/} \cite{suth}.
This model exhibits $P-1$ gapless elementary
excitation branches.  We presume, based on what happens in the SU(2)
model \cite{fata}, that for a chain of $N$ sites where $N$ is an
integer multiple of $P$, the ground state is an SU($P$) singlet and the
low-lying excitations transform according either to the singlet or
the adjoint representation. This is essentially what happens in the
fermion mean field (large $P$) theory of the
SU($P$) antiferromagnet \cite{bza,AA,RS}. The mean field has four degenerate
quarter-filled ($\kF=\fourth\pi$) bands for $P=4$.  Although there it
has no true long-ranged order, the spin and isospin susceptibilities
diverges at the nesting wavevector $2\kF=\half\pi$, which describes a
commensurate spin density wave of period four.  The period four arises
because the spin chain is in its fundamental representation, and by
`4-ality' one needs four sites to make a singlet \cite{comm3}. The mean
field theory also predicts a constant uniform (Pauli) susceptibility, and
a linear specific heat as in a Fermi liquid\cite{AA}.

{\em Ferro-Antiferromagnetic points --}
Along the surface $\ubar_2=\ubar_0$, our Hamiltonian possesses an
SU(2)$\times$SU(2) symmetry.  There are then two special limits in
which we can determine the exact ground state.
(i) The ``$\rmF\times\rmA$ model''
at $\ubar_0\to\infty$, with $J_\parallel= 2 (\tpar)^2/\ubar_1$,
\be
\cH_{\rmF\times\rmA}=-{4(\tpar)^2\over\ubar_1}\sum_n
(\bfS_n\cdot\bfS_{n+1}+\frac{3}{4})
(\fourth - \bfI_n\cdot\bfI_{n+1})\label{HamFA}
\ee
where the interactions are ferromagnetic in
the spin channel and antiferromagnetic in the isospin channel, and
(ii) the ``$\rmA\times\rmF$  model'' for
$\ubar_1\to\infty$, with $J_\parallel= 2 (\tpar)^2/ \ubar_0$, and the
roles of $\bfI$ and $\bfS$ interchanged.

It is possible to prove that the ground state of $\cH_{\rmF\times\rmA }$
is the fully polarized ferromagnet $\ket{\rmF}_S$ for the spin variables,
and Bethe's ground of the spin-half antiferromagnet for the
isospin variables \ie
\be
\Psi^{\rmF\times\rmA }_0 =  \ket{\rmF}_S\otimes\ket{{\rm Bethe}}_I\ .
\label{prodwf}
\ee
A corresponding result holds for $\cH_{\rmA\times\rmF}$, with spin and isospin
variables exchanged.
Due to the $SU(2)\otimes SU(2)$ symmetry the total spin $S_{\rm tot}$, total
isospin $I_{\rm tot}$,
and their polarizations along the $\zhat$ axis ($M_S$ and $M_I$,
respectively) are good quantum numbers.
Following Lieb and Mattis' proof of the Marshall theorem for the
Heisenberg model\cite{LM}, we perform  a $\pi$ rotation about the
$\zhat$ axis of the isospin operators on odd-numbered sites.
The Hamiltonian transforms into  a non-positive (`negative semidefinite')
operator in the product Ising basis
\bea
\cH_{\rmF\times\rmA }&\to&
J\sum_n(I^z_n I^z_{n+1}-\half I^+_n I^-_{n+1}-\half I^-_n I^+_{n+1}
-\fourth)\nonumber\\
&&\qquad\times
(\bfS_n \cdot \bfS_{n+1} +\frac{3}{4})\equiv\cH'_{\rmF\times\rmA }
\eea
The accessibilty of all states within a given
magnetization sector by repeated application of the Hamiltonian,
implies (see Ref. \cite{LM}) that the ground state of
$\cH'_{\rmF\times\rmA }$ in the sector
$(M_S,M_I)=(0,0)$ can be chosen to be positive definite in the
sublattice-rotated Ising basis, \ie\ it obeys Marshall's sign rule.
Since the same Marshall signs hold for the state on the right hand side
of  Eq. \ref{prodwf},
which has $S_{\rm tot}=\half N $, and $I_{\rm tot}=0$,
the two sides of Eq. \ref{prodwf} have finite overlap hence the same
$S_{\rm tot}$ and $I_{\rm tot}$.
We are free to choose $M_S=\half N$ as a representative of the ground
state manifold.
Note that $\ket{\Psi^{\rmF\times\rmA  }_0}$ is indeed an eigenstate of the
spin triplet
projection operator $(\bfS_n\cdot\bfS_{n+1}+\frac{3}{4})$ with
eigenvalue one.
It follows from Eq. \ref{HamFA} that the isospin part of the wavefunction
is the ground state of the spin-half
antiferromagnetic Heisenberg chain, given by Bethe's {\it Ansatz}.

Exact excitations of $\cH_{\rmF\times\rmA}$
within the isospin sector (retaining full spin
polarization) with dispersion $\half\pi J|\sin k|$ can be constructed
as in Refs. \cite{fata,dcp}.  The gapless ferromagnetic magnons,
which exist due to Goldstone's theorem, can be approximated within the
Single Mode Approximation (SMA): $\ket{k}\equiv
S^-_k\ket{\Psi_0^{\rmF\times\rmA}}$.  The trial state dispersion is
\be
\omega( k) \le  2\ln( 2)J
 (1-\cos k )
\ee
from which we see that the ferromagnon bandwidth is decreased due to the
antiferromagnetic nearest-neighbor isospin correlations, \ie\
$\langle \fourth - \bfI_n \cdot \bfI_{n+1} \rangle = \ln(2)$.

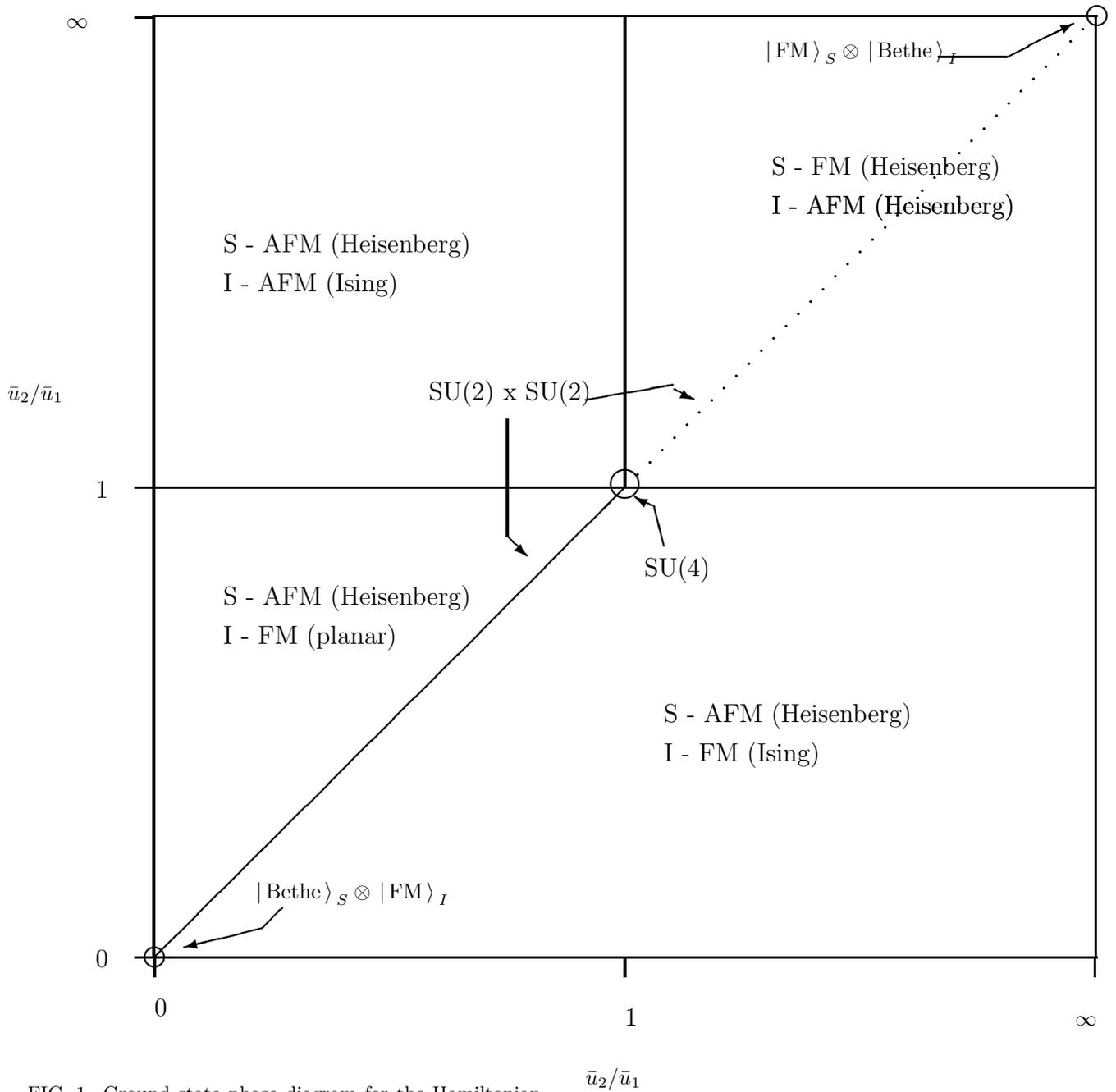
\begin{figure}

\setlength{\unitlength}{0.0125in}%
\begin{picture}(561,595)(5,215)
\thicklines
\put( 80,280){\circle{10}}
\put(561,761){\circle{10}}
\put( 80,280){\framebox(480,480){}}
\put( 80,280){\line( 1, 1){240}}
\multiput(320,520)(6.31579,6.31579){39}{\makebox(0.4444,0.6667){\tenrm .}}
\put( 80,520){\line( 1, 0){480}}
\put(320,760){\line( 0,-1){240}}
\multiput(320,520)(6.31579,6.31579){39}{\makebox(0.4444,0.6667){\tenrm .}}
\put(340,490){\line(-1, 4){  5}}
\multiput(335,510)(-0.50000,0.25000){21}{\makebox(0.4444,0.6667){\sevrm .}}
\put(325,515){\vector(-2, 1){0}}
\put(300,565){\line( 6, 1){ 44.595}}
\multiput(345,570)(0.50000,-0.25000){21}{\makebox(0.4444,0.6667){\sevrm .}}
\put(355,565){\vector( 2,-1){0}}
\put(260,555){\line( 0,-1){ 60}}
\multiput(260,495)(0.40000,-0.40000){26}{\makebox(0.4444,0.6667){\sevrm .}}
\put(270,485){\vector( 1,-1){0}}
\multiput(145,305)(-0.40000,-0.40000){26}{\makebox(0.4444,0.6667){\sevrm .}}
\put(135,295){\vector(-4,-1){ 40}}
\put(480,740){\line( 1, 0){ 35}}
\put(515,740){\line( 0, 1){  0}}
\put(515,740){\vector( 2, 1){ 34}}
\put(320,280){\line( 0,-1){  5}}
\put(320,275){\line( 0,-1){  5}}
\put(560,280){\line( 0,-1){ 10}}
\put( 80,280){\line( 0,-1){ 10}}
\put( 80,280){\line(-1, 0){ 10}}
\put( 80,520){\line(-1, 0){ 10}}
\put( 80,760){\line(-1, 0){ 10}}
\put(115,460){\makebox(0,0)[lb]{\raisebox{0pt}[0pt][0pt]{\twlrm S - AFM
(Heisenberg)}}}
\put(320,522){\circle{14}}
\put(340,400){\makebox(0,0)[lb]{\raisebox{0pt}[0pt][0pt]{\twlrm S - AFM
(Heisenberg)}}}
\put(250,795){\makebox(0,0)[lb]{\raisebox{0pt}[0pt][0pt]{\twlrm }}}
\put(115,640){\makebox(0,0)[lb]{\raisebox{0pt}[0pt][0pt]{\twlrm S - AFM
(Heisenberg)}}}
\put(395,680){\makebox(0,0)[lb]{\raisebox{0pt}[0pt][0pt]{\twlrm S - FM
(Heisenberg)}}}
\put(395,660){\makebox(0,0)[lb]{\raisebox{0pt}[0pt][0pt]{\twlrm I - AFM
(Heisenberg)}}}
\put(395,660){\makebox(0,0)[lb]{\raisebox{0pt}[0pt][0pt]{\twlrm I - AFM
(Heisenberg)}}}
\put(115,620){\makebox(0,0)[lb]{\raisebox{0pt}[0pt][0pt]{\twlrm I - AFM
(Ising)}}}
\put(115,440){\makebox(0,0)[lb]{\raisebox{0pt}[0pt][0pt]{\twlrm I - FM
(planar)}}}
\put(340,380){\makebox(0,0)[lb]{\raisebox{0pt}[0pt][0pt]{\twlrm I - FM
(Ising)}}}
\put(330,475){\makebox(0,0)[lb]{\raisebox{0pt}[0pt][0pt]{\twlrm SU(4)}}}
\put(220,565){\makebox(0,0)[lb]{\raisebox{0pt}[0pt][0pt]{\twlrm SU(2) x
SU(2)}}}
\put(130,310){\makebox(0,0)[lb]{\raisebox{0pt}[0pt][0pt]{\twlrm $\ket{\rm
Bethe}_S\otimes\ket{{\rm FM}}_I$ }}}
\put(390,740){\makebox(0,0)[lb]{\raisebox{0pt}[0pt][0pt]{\twlrm $\ket{{\rm
FM}}_S\otimes\ket{{\rm Bethe} }_I$}}}
\put( 80,250){\makebox(0,0)[lb]{\raisebox{0pt}[0pt][0pt]{\twlrm 0}}}
\put(320,245){\makebox(0,0)[lb]{\raisebox{0pt}[0pt][0pt]{\twlrm 1}}}
\put(550,245){\makebox(0,0)[lb]{\raisebox{0pt}[0pt][0pt]{\twlrm $\infty$}}}
\put( 50,275){\makebox(0,0)[lb]{\raisebox{0pt}[0pt][0pt]{\twlrm 0}}}
\put( 50,515){\makebox(0,0)[lb]{\raisebox{0pt}[0pt][0pt]{\twlrm 1}}}
\put( 35,755){\makebox(0,0)[lb]{\raisebox{0pt}[0pt][0pt]{\twlrm $\infty$}}}
\put(  5,565){\makebox(0,0)[lb]{\raisebox{0pt}[0pt][0pt]{\twlrm ${\bar
u}_2/{\bar u}_1$}}}
\put(300,215){\makebox(0,0)[lb]{\raisebox{0pt}[0pt][0pt]{\twlrm ${\bar
u}_2/{\bar u}_1$}}}
\end{picture}

\caption{
Ground state phase diagram for the Hamiltonian $\cHt$ of Eq. \protect
\ref{sex1}.
}
\label{fig-phase}
\end{figure}

{\em Classical Phase Diagram.}
The ground state depends on the dimensionless  ratios
${\bar u}_0/{\bar u_1}$ and  ${\bar u}_2/{\bar u_1}$. The classical
approximation (justified at  $S,I>>1$) is given by minimizing the bond
energies of Eq. (\ref{sex1}) as function of vectors  $\bfS_\bfi $ and
$\bfI_\bfi$ of magnitude $\half$.  The results are plotted in
Fig. \ref{fig-phase}).

It is interesting to note that the
SU(4) symmetry point is at the border of 4 distinct ordered phases of
different symmetries, where the energy is degenerate along the lines
$\langle \bfI_n\cdot\bfI_{n+1}\rangle=-{1\over 4}$ and
$\langle \bfS_n\cdot\bfS_{n+1}\rangle={1\over 4}$.
The large degeneracy of the classical SU(4) model is reduced
by quantum fluctuations.

The classical regime of Heisenberg spin-ferromagnetism and
isospin-antiferromagnetism extends throughout $\ubar_0,\ubar_2 > \ubar_1$,
although quantum fluctuations break the SU(2) isospin symmetry away from
the isotropic line $\ubar_0=\ubar_2$ (marked as a dashed line in
Fig. \ref{fig-phase}).

{\em 3D Ordering in the $\rmF\times\rmA$ Model --}
As shown by Scalapino \etal \cite{sip}, one can treat
the interchain interactions by mean field theory and thereby derive an
expression for the full susceptibility $\xhi_{ab}(\bfq_\perp,q_z,\omega)$
in terms of $\xhi^{\rm 1D}_{ab}(q_z,\omega)$, the susceptibility for
the one-dimensional chains.  The general result is
\be
\xhi(\bfq_\perp,q_z,\omega)=\sbl \mbox{\bf 1}-J_\perp(\bfq_\perp)\,\xhi^{\rm
1D}
(q_z,\omega)\sbr^{-1}\xhi^{1D}(q_z,\omega)\ ,
\label{susc}
\ee
where $J_\perp(\bfq_\perp)=\sum_{\bdper} J_\perp(\bdper)\,
e^{-i\bfq_\perp\cdot\bdper}$
is the spatial Fourier transform of the interchain
coupling matrix.  (Note that the quantities $\xhi$, $J_\perp$, and
$\xhi^{\rm 1D}$ in Eq. \ref{susc} are matrices.)  This approximation
also may be employed at finite temperature.

Consider now the F$\times$A model discussed above.  At finite
temperature $T$, long-ranged ferromagnetic order is destroyed and
the global SU(2)$\times$SU(2) symmetry is restored.
The uniform susceptibility of the ferromagnetic chain is given by
\be
\xhi_\rmF'(0,0;T)={J_\parallel\over 24 T^2}+\ldots\ ,
\ee
as was first derived by Takahashi in Ref. \cite{tak}
(see also Refs. \cite{AA,assa}).

For the antiferromagnetic susceptibility, we appeal to the bosonization
results of Schulz and of Eggert and Affleck \cite{ian2}, who have
computed the dynamic susceptibility of the $S=\half$ antiferromagnetic
Heisenberg
chain. Performing a Fourier transform of their result and taking the
low frequency and wavevector limit near the antiferromagnetic point
we obtain the {\it staggered} isospin susceptibility
\be
\xhi_\rmA  \approx {a_0^2\over\pi T}
\ee
where $a_0\simeq 4.44$.

For mixed interchain coupling operators
\eg\  $\cO=S^x I^y$, we may use the assumed independence of
ferromagnetic and antiferromagnetic magnons to obtain
at low
temperatures $\xhi'_{\rmF\rmA}(\pi,0;T)\sim (J_\parallel T)^{-1/2}$,
which diverges even more slowly than $\xhi_\rmA$ in the $T\to 0$ limit.

The interchain interaction is given by
$J_\perp=J_\parallel((t^\perp_1)^2+(t^\perp_2)^2)/4(t^\parallel)^2$,
where $t^\perp_{1},t^\perp_{2}$
are the transverse hopping integrals (see Ref. \cite{long}).
Thus, as the temperature is lowered, a transition from paramagnetic to
ferromagnetic state should set in when $J_\perp\xhi'_\rmF=1$.
This yields $T_\rmC\simeq\sqrt{J_\parallel J_\perp/24}$.  The relation
$T_\rmC\propto\sqrt{J_\parallel J_\perp}$ was also found by
Scalapino \etal\ (ref. \cite{sip}) in their studies of anisotropic
Heisenberg magnets.  It is conceivable that at still
lower temperatures a N{\'e}el ordering of the isospin variables occurs at
a N{\'e}el temperature $T_\rmN\simeq 3 a_0^2 J_\perp/\pi$.

{\em Experimental notes}:
(a) The lower isospin
transition, to our knowledge, has not been resolved experimentally. Perhaps
it is not very well separated from the ferromagnetic transition which
would help explain the mysterious excessive entropy of transition found by
Ref.\cite{tanaka}.

(b) Alternatively, the isospin ordering might be preempted by a
{\em isospin-Peierls}
ordering (orbital dimerization) aided by the electron-phonon coupling. In
that case, a signature for the isospin-Peierls effect should be present
in X-ray scattering or in the phonon spectrum.

(c) The role of a possible
orientational disordered ground state \cite{brooks} has not been considered
here
although it might help explain the observed weak ferromagnetism
\cite{tdae-fm}. In addition,
Bloch's $T^{3/2}$ temperature dependence of the ordered moment found in
Ref. \cite{tdae-fm1} which holds upto $T\approx T_\rmC$ is hard to reconcile
with  quasi-one dimensionality where $J_\perp << T_\rmC$.

In summary,
we have introduced a new model of
quasi-one dimensional interacting electrons with doubly degenerate orbitals
motivated by the structure of TDAE-\c60. At occupancy of one
electron per site, we obtain a Mott-insulator with  muticomponent
superexchange between
spins and isospins at neighboring sites. At special values of the
interactions we identify exactly solvable points, including the SU(4)
antiferromagnet, and spin-ferromagnet, isospin-antiferromagnet limit.
The classical ground state diagram also
contains a large region of spin ferromagnetism and orbital
antiferromagnetism which we believe is relevant for  TDAE-\c60.
A mean field analysis of the interchain coupling in this regime
predicts two transition temperatures: ferromagnetic spin ordering at
$T_\rmC\propto \sqrt{J_\parallel J_\perp}$, and  orbital (isospin)
antiferromagnetic ordering at $T_\rmN\propto J_\perp $. This lower
transition, to our knowledge, has not yet been resolved experimentally.

{\em Acknowledgements}
We thank Ian Affleck, Amnon Aharony, Brooks Harris, and Ganpathy Murthy
for useful discussions.
The support of US-Israeli Binational Science Foundation, the Israeli Academy
of Sciences and the Fund for Promotion of Research at Technion is gratefully
acknowledged.  DPA thanks the Institute for Theoretical Physics at
Technion, where this work was performed, and the Lady Davis Fellowship
Trust for partial support.

\vfill\eject


\begin{references}
\bibitem{comm1} Other examples are:  the Nagaoka
ferromagnet of one hole in the infinite $U$
Hubbard model, and the Hubbard model on a bipartite lattice with unequal
sublattice sizes.  Neither is easy to realize physically.
\end{references}
\end{document}